\documentclass[aps,preprint]{revtex4}%
\usepackage{amssymb}
\usepackage{amsfonts}
\usepackage{amsmath}
\usepackage{graphicx}%
\providecommand{\U}[1]{\protect\rule{.1in}{.1in}}

\begin{document}
\title{{\Large Kidnapping Model: An Extension of Selten's Game}}
\author{Azhar Iqbal$^{\dagger}$, Virginie Masson$^{\ddagger}$, and Derek
Abbott$^{\dagger}$}
\affiliation{$^{\dagger}${\small School of Electrical \& Electronic Engineering, University
of Adelaide, South Australia 5005, Australia}}
\affiliation{$^{\ddagger}${\small School of Economics, University of Adelaide, South
Australia 5005, Australia}.}

\begin{abstract}
Selten's game is a kidnapping model where the probability of capturing the
kidnapper is independent of whether the hostage has been released or executed.
Most often, in view of the elevated sensitivities involved, authorities put
greater effort and resources into capturing the kidnapper if the hostage has
been executed, in contrast to the case when a ransom is paid to secure the
hostage's release. In this paper, we study the \textit{asymmetric game} when
the probability of capturing the kidnapper depends on whether the hostage has
been executed or not and find a new uniquely determined perfect equilibrium
point in Selten's game.

\end{abstract}
\maketitle

\section{Introduction}

Although it is not a common crime, there are parts of the world where
kidnapping is a real and constant threat. In a typical scenario, a victim is
abducted and a monetary demand is made for his or her release. Although it
appears to be a simple exchange of money for the release of a hostage, dealing
with the situation does require considerable planning. This is where game
theory \cite{Rasmusen,Osborne,Binmore}, a developed branch of mathematics that
models strategic situations, can offer valuable insights.


In 1976, Reinhard Selten \cite{Selten0} developed a game-theoretic model of
kidnapping as a two-person sequential game between player K (Kidnapper) and
player F (hostage's Family). The game begins with K's choice whether or not to
go ahead with his plan that is described by a binary decision variable $b:$%

\begin{equation}
b:\left\{
\begin{array}
[c]{c}%
0\\
1
\end{array}
\right.
\begin{array}
[c]{c}%
\text{Not to kidnap,}\\
\text{To kidnap.}%
\end{array}
\label{b_Def}%
\end{equation}

The game ends if K selects $b=0$. If K selects $b=1$, he takes the hostage to
a hidden place unknown to player F and to the police, and announces a ransom
demand $D$.

Numerous questions then arise. Will the hostage's family pay the ransom $D$ or
will they try to negotiate a lower amount? If they do pay the ransom, should K
free the hostage instead of executing him/her? Moreover, if F does not expect
K to free the hostage, why should it be expected that F pay some ransom?

It is assumed that, on knowing the demand $D$, a negotiation process starts
between players K and F. Player F makes an offer $C$ which is the amount
willing to be paid, and player K either decides to accept $C$ and release the
hostage, or to execute the hostage. The situation can be seen as a simple
description of an extended bargaining process.

In his model, Selten assumed that K's threat to execute the hostage has some
credibility, even though K cannot improve his situation by doing so. In
particular, it is expected that with a positive probability $\alpha$, player K
may deem the offer $C<D$ to be insufficient, and thus decide to execute the
hostage. Selten assumed that the probability $\alpha$ can be described as a
linear function of $\frac{C}{D}:$%

\begin{equation}
\alpha=a\left[  1-\frac{C}{D}\right]  \text{ for }0\leq C\leq D \label{Alpha}%
\end{equation}
where $a$ is a constant with $0<a<1$. The parameter $\alpha$ thus describes
K's non-rational decision to execute the hostage, as in this case, the
traditional utility maximisation principle is ignored.

It is nonetheless possible that Player K makes the rational decision to
execute the hostage independently of whether the offer $C$ is deemed
insufficient or not. Selten used a binary decision variable $e$ to describe
this situation:%

\begin{equation}
e:\left\{
\begin{array}
[c]{c}%
0\\
1
\end{array}
\right.
\begin{array}
[c]{c}%
\text{Release of hostage for ransom }C\text{,}\\
\text{Execution of hostage,}%
\end{array}
\label{Def_e}%
\end{equation}
i.e. even if an offer is made at $C$, $0\leq C\leq D$, the hostage can still
be executed for some $C$.

\section{Modified kidnapping game}

In either case of the hostage having been executed or released, the
authorities will put efforts in finding and capturing the kidnapper K. In
Selten's model, it is assumed that, in both cases, the authorities will be
successful in capturing K with some probability $q$, where%

\begin{equation}
0<q<1 \label{q}%
\end{equation}
i.e.~the probability of capturing the kidnapper is independent of whether the
hostage has been released or executed. We ask whether this indeed is the usual case.

Cursory observations of media coverage related to kidnapping incidences
highlight the political pressure faced by authorities to severely punish those
responsible, the idea being that punishment helps decrease the incidence of
such events in the future. It is however unclear whether authorities favour
the allocation of extra resources towards the capture of those who executed
the hostage or those who did not. No consensus appears to prevail, and
resource spending seems to be case dependent and government dependent, with
the media likely playing a role in the decision of whether extra resources are
spent to increase the chances of capturing the kidnapper.

Assuming that increased spending leads to a higher probability of capture, we
thus adopt two random choice parameters $q_{0}$ and $q_{1}$ instead of the
fixed probability $q$ assumed by Selten, that we define as follows:%
\begin{align}
q_{0}  &  =\text{probability of capture of K if hostage is released,}%
\nonumber\\
q_{1}  &  =\text{probability of capture of K if hostage is killed.}
\label{Probs_Def}%
\end{align}

This allows us to an improved modeling of the responses by authorities, and
helps us elucidate whether allocating extra resources to increase the
likelihood of capturing the kidnapper influences the kidnapper's strategy.

As we focus on the probability of the kidnapper's capture, it is also
important to encompass the idea that, in the case where the hostage is
executed, families derive a higher disutility from the kidnapper still being
at large. This gives us the payoffs depicted in Table 1.

Elevated sensitivities lead to increased pressure on the police, and
government, to capture the kidnappers that usually results in an increase in
the resources for finding the kidnappers. Investing extra efforts and
resources may result in an increased probability of capturing the kidnapper.

In the following, we study the situation when the probability of capturing the
kidnapper depends on whether the hostage has been killed or not and find a new
uniquely determined perfect equilibrium point in Selten's game.%

\begin{gather*}
\text{
\ \ \ \ \ \ \ \ \ \ \ \ \ \ \ \ \ \ \ \ \ \ \ \ \ \ \ \ \ \ \ \ \ \ \ \ \ \ \ \ \ \ \ \ \ \ \ \ \ \ \ \ \ \ \ \ \ \ \ \ \ \ \ \ \ \ \ \ \ \ \ \ \ \ \ \ \ \ \ \ \ \ \ \ \ \ \ \ \ \ }%
\begin{array}
[c]{c}%
\text{Payoffs}%
\end{array}
\\%
\begin{tabular}
[c]{lrr}%
$\text{Outcome}$ & Player K & Player F\\
$\text{Kidnapping does not take place}$ & \multicolumn{1}{c}{$0$} &
\multicolumn{1}{c}{$0$}\\
Release of hostage for ransom payment $C$, Kidnapper not caught &
\multicolumn{1}{c}{$C$} & \multicolumn{1}{c}{$-C$}\\
$\text{Kidnapper caught after release of hostage}$ & \multicolumn{1}{c}{$-X$}
& \multicolumn{1}{c}{$0$}\\
$\text{Kidnapper not caught after execution of hostage}$ &
\multicolumn{1}{c}{$-Y$} & \multicolumn{1}{c}{$-W_{1}$}\\
$\text{Kidnapper caught after execution of hostage}$ & \multicolumn{1}{c}{$-Z$%
} & \multicolumn{1}{c}{$-W_{2}$}%
\end{tabular}
\\
\\
\text{Table 1. The payoffs for players K and F.}%
\end{gather*}

Here $W_{1},$ $W_{2},$ $X,$ $Y,$ and $Z$ are positive constants and utilities
of K and F that are assumed to be linear in money. In the original game
presented by Selten, $W_{1}=W_{2}=W$, and $q_{0}=q_{1}=q$.

As in Selten's game, if K is caught, the execution of the hostage results in
an increased disutility relative to the case when K releases the hostage. Thus,%

\begin{equation}
Z\geq X.
\end{equation}
As the complete history of the previous game is known to both players at every
point in the course of play, the game can be identified as an extensive game
with perfect information \cite{Selten1}.

Note that Table 1 encompasses a number of simplifying assumptions. First,
player K's cost of preparing the kidnapping is assigned zero value. Also,
player F's non-monetary disutilities, other than those that are incurred if
player K executes the hostage, are ignored. In reality, there can be
significant disutility for player F resulting from the emotional stress of
engaging in a bargaining process with player K. Also, player F is assumed to
gain no utility from the capture of K by the authorities if the hostage is
released. Finally, note that utilities when player K is caught after the
release of the hostage do not depend on the ransom money $C$ as it is
recovered and given back to player F. However, player K is then left with
disutility $X$. FIG. 1 shows the extensive form of the game.%

\begin{figure}[ptb]%
\centering
\includegraphics[
height=5.1811in,
width=6.4835in
]%
{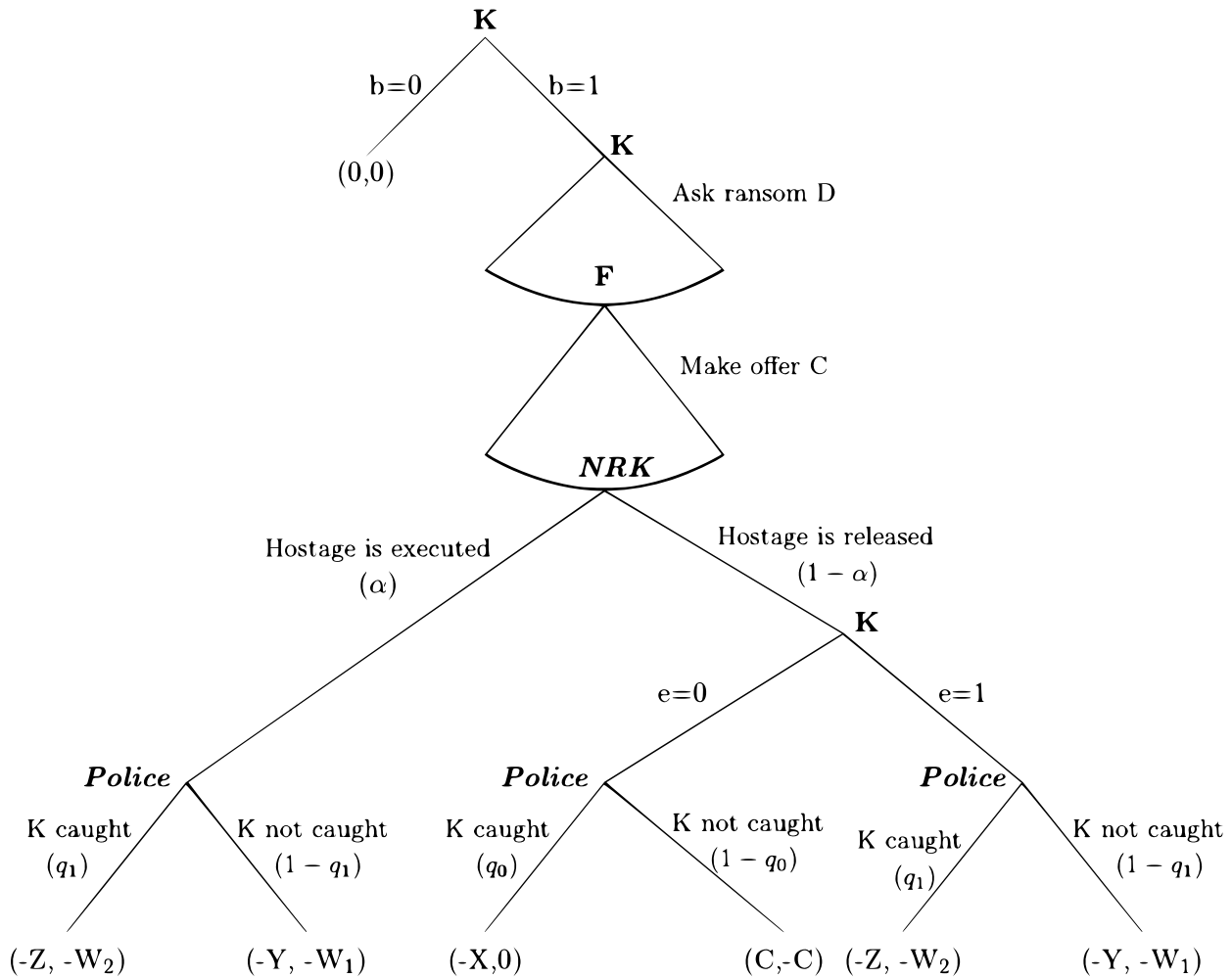}%
\caption{Representation of the game in extensive form. Strategic players, i.e.
those deriving payoffs from the game, are K (Kidnapper) and F (hostage's
Family). NRK represents the nonrational decision of player K. NRK and Police's
later actions represent random events occurring with some probabilities.
Payoffs are presented for player K first and player F second.}%
\end{figure}

\subsection{Game timeline}

Below, we provide a short description of the timeline of the game:

\begin{enumerate}
\item Player K chooses between $b=1$ and $b=0$, i.e. whether or not to kidnap someone.

\item If K selects $b=0$, the game ends and both players K and F receive zero
payoffs; if K selects $b=1$, K then announces demand $D$.

\item After observing demand $D$, F makes an offer $C$ such that $0\leq C\leq
D$.

\item After K observes the offer $C$, a non-rational execution of the hostage
occurs with probability $\alpha$ defined by Eq.~(\ref{Alpha}).

\item If the hostage is not executed non-rationally, K chooses between $e=0$
and $e=1$. If K selects $e=0$, this means that ransom $C$ is paid and the
hostage is released. If K selects $e=1$, the hostage is (rationally) executed
irrespective of whether any ransom is paid or not.

\item After release or execution of the hostage, two final random choice
parameters $q_{0}$ and $q_{1}$ reflect the likelihood for the kidnapper to be
captured, where $q_{0}$ is the probability of capture if the hostage is
released and $q_{1}$ is the probability of capture if the hostage is executed.

\item The game then ends with payoffs according to Table 1.
\end{enumerate}

\section{Equilibrium of the game}

Being one of the foundational concepts in game theory, Nash equilibrium are
used to predict the strategies used by players of noncooperative games. A
strategy profile specifies a strategy for each player and constitutes a Nash
equilibrium if no player has an incentive to deviate from their current
strategy. Any finite game admits at least one Nash equilibrium. The
mathematical conditions defining a Nash equilibrium, called the Nash
conditions, may nonetheless lead to unreasonable outcomes, as pointed out by
Selten \cite{Selten,Selten1}. This is because the Nash conditions do not
account for the dynamics of the game (if any).

Selten thus used the notion of \textit{perfect equilibrium} as a refinement on
the set of Nash equilibria to solve for the equilibrium of the Kidnapping
game. A \textit{subgame perfect equilibrium}
\cite{Selten,Selten1,Selten2,Kalai} is not only a Nash equilibrium in the
whole game, it is also a Nash equilibrium is every subgame. For finite games
with perfect information, such as the one considered in this paper, subgame
perfect equilibrium are commonly determined using backward induction
\cite{Osborne,Rasmusen}.

In what follows, we follow Selten's original work, and identify the
equilibrium of the game using the concept of subgame perfection
\cite{Kuhn,Osborne,Binmore1}.

\subsection{Optimal choice of $e$}

We start by examining the subgame that begins with player K's choice of $e$.
Let $V_{0}$ be K's expected payoff if K selects $e=0$ and let $V_{1}$ be the
expected payoff if K selects $e=1$ (i.e.~the execution of the hostage). We have:%

\begin{equation}
V_{0}=(1-q_{0})C-q_{0}X,\text{ and}. \label{Opt_Choice_e_eq1}%
\end{equation}

\begin{equation}
V_{1}=-(1-q_{1})Y-q_{1}Z. \label{Opt_Choice_e_eq2}%
\end{equation}

Note that the case%

\begin{equation}
q_{0}=q_{1}=q \label{Selten_case}%
\end{equation}
was considered by Selten.

In that case, as $C\geq0,$ $Y>0$, $Z\geq X$ and $0<q<1$ we have%

\begin{equation}
V_{0}>V_{1}%
\end{equation}
and $e=0$ becomes the optimal choice for player K. That is, for the case
studied by Selten, player K will never rationally decide to execute the
hostage. This could give the impression that when (\ref{Selten_case}) does not
hold, release of the hostage $(e=0)$ would not remain the optimal choice of
$e$. \vspace{0pt} If the values of $q_{0}$ and $q_{1}$ rely on heightened
sensitivities, i.e.~authorities allocate more resources when the hostage has
been executed so as to increase the likelihood of capturing K, we can assume
that $q_{1}>q_{0}$. In this case, $e=0$ remains the optimal choice for K, as
in Selten's original work. This is because $(1-q_{0})C>0$, while
$-(1-q_{1})Y<0$, and thus disutilities are ranked such that $q_{0}X<q_{1}Z$
$($as $X\leq Z)$, which gives us that $V_{0}>V_{1}$.

However, if $q_{0}>q_{1}$, release of the hostage $(e=0)$ may \textit{not}
remain the optimal choice for K, which is in contrast with Selten's work.

\subsection{Optimal choice of $C$}

In the subgame that begins with player F's choice of $C$, player F knows that
player K can execute the hostage with probability $\alpha$ given in
Eq.~(\ref{Alpha}). Using Table 1, the expected value of F's utility is thus
equal to:%

\begin{equation}
U=(1-\alpha)\left[  (1-q_{0})(-C)+q_{0}(0)\right]  +\alpha\left[
(1-q_{1})(-W_{1})+q_{1}(-W_{2})\right]  . \label{U_New}%
\end{equation}

With the constraints%

\begin{equation}
W_{1}=W_{2}=W\text{ and }q_{0}=q_{1}=q, \label{Constraints}%
\end{equation}
Eq.~(\ref{U_New}) is reduced to%

\begin{equation}
U=-(1-\alpha)(1-q)C-\alpha W, \label{U_Selten}%
\end{equation}
which is player F's expected value of utility for the case that Selten
considered. From FIG. 1 we note that $(1-\alpha)$ is the probability of the
hostage being released because of K's non-rational decision. However, $e=0$ is
K's rational decision to release the hostage and thus $(1-\alpha)$ is not the
probability of $e=0$.

Using the expression for $\alpha$ from Eq.~(\ref{Alpha}) in Eq.~(\ref{U_New}),
we have:%

\begin{align}
U  &  =\left[  -a(1-q_{0})\right]  \frac{C^{2}}{D}+\left[  (1-q_{1}%
)W_{1}+q_{1}W_{2}\right]  \frac{aC}{D}-\nonumber\\
&  (1-a)(1-q_{0})C-\left[  (1-q_{1})W_{1}+q_{1}W_{2}\right]  a. \label{U_New1}%
\end{align}

Once again, under the constraints described by (\ref{Constraints})
Eq.~(\ref{U_New1}) reduces to%

\begin{equation}
U=-a(1-q)\frac{C^{2}}{D}+\left[  \frac{aW}{D}-(1-a)(1-q)\right]  C-aW,
\end{equation}
which is a strictly concave quadratic function as obtained by Selten
\cite{Selten0}. In order to determine the optimal value $\bar{C}$ of $C$ we
compute $\frac{\partial U}{\partial C}$ from Eq.~(\ref{U_New1})%

\begin{equation}
\frac{\partial U}{\partial C}=\left[  -a(1-q_{0})\right]  \frac{2C}{D}+\left[
(1-q_{1})W_{1}+q_{1}W_{2}\right]  \frac{a}{D}-(1-a)(1-q_{0}). \label{U_P}%
\end{equation}
Eq.~(\ref{U_P}) shows that $U$ assumes its maximum at%

\begin{equation}
C^{\prime}=\frac{(1-q_{1})W_{1}+q_{1}W_{2}}{2(1-q_{0})}-\frac{(1-a)D}{2a},
\label{C_New}%
\end{equation}
if the value of $C$ is in the interval $0\leq C\leq D$. This is the case if
$D$ is in the closed interval between the following critical values%

\begin{align}
D_{1}^{\prime}  &  =\frac{a}{(1+a)}.\frac{(1-q_{1})W_{1}+q_{1}W_{2}}%
{(1-q_{0})},\label{D1_New}\\
D_{2}^{\prime}  &  =\frac{a}{(1-a)}.\frac{(1-q_{1})W_{1}+q_{1}W_{2}}%
{(1-q_{0})}. \label{D2_New}%
\end{align}

As before, under the constraints (\ref{Constraints}) Eqs.~(\ref{D1_New}%
,~\ref{D2_New}) become%

\begin{align}
D_{1}  &  =\frac{a}{(1+a)}.\frac{W}{(1-q)},\label{D1}\\
D_{2}  &  =\frac{a}{(1-a)}.\frac{W}{(1-q)}, \label{D2}%
\end{align}
as obtained by Selten. To determine the range for which $U$ is an increasing
function i.e.%

\begin{equation}
\frac{\partial U}{\partial C}>0, \label{inequality_1}%
\end{equation}
we use Eqs.~(\ref{U_P},~\ref{D1_New}) to write inequality (\ref{inequality_1}) as%

\begin{equation}
\left[  (1+a)D_{1}^{\prime}-(1-a)D\right]  >2aC,
\end{equation}
i.e.~the function $U,$ as described by Eq.~(\ref{U_New1}), is an increasing
function for $D_{1}^{\prime}>D.$ Likewise, considering the inequality%

\begin{equation}
\frac{\partial U}{\partial C}<0, \label{inequlity_2}%
\end{equation}
we use Eqs.~(\ref{U_P},~\ref{D2_New}) to write inequality (\ref{inequlity_2}) as%

\begin{equation}
(D_{2}^{\prime}-D)(1-a)<2aC,
\end{equation}
i.e.~the function $U,$ as described by Eq.~(\ref{U_New1}), is a decreasing
function for $D_{2}^{\prime}<D.$

In view of Eq.~(\ref{C_New}) describing the maximum that the function $U$
assumes, player F's optimal offer $\bar{C}^{\prime}$ can be described as follows%

\begin{equation}
\bar{C}^{\prime}=\left\{
\begin{array}
[c]{c}%
D\\
\frac{(1-q_{1})W_{1}+q_{1}W_{2}}{2(1-q_{0})}-\frac{(1-a)D}{2a}\\
0
\end{array}
\right.
\begin{array}
[c]{l}%
\text{for }0<D\leq D_{1}^{\prime},\\
\text{for }D_{1}^{\prime}<D\leq D_{2}^{\prime},\\
\text{for }D>D_{2}^{\prime}.
\end{array}
\label{Opt_Offer_New}%
\end{equation}
As $D$ increases, the optimal offer $\bar{C}^{\prime}$ first increases up to
$D_{1}^{\prime}$ and then decreases until it becomes $0$ at $D=D_{2}^{\prime}%
$. In the interval $D_{1}^{\prime}<D\leq D_{2}^{\prime}$ the optimal offer
$\bar{C}^{\prime}$ is decreased by an increase of $D$. The threat of execution
of the hostage is avoided in the intervals $0<D\leq D_{1}^{\prime}$, as player
F agrees to meeting the demand for the ransom.

Note that under constraints (\ref{Constraints}), the optimal offer $\bar
{C}^{\prime}$ is reduced to $\bar{C}$%

\begin{equation}
\bar{C}=\left\{
\begin{array}
[c]{c}%
D\\
\frac{W}{2(1-q)}-\frac{(1-a)D}{2a}\\
0
\end{array}
\right.
\begin{array}
[c]{l}%
\text{for }0<D\leq D_{1},\\
\text{for }D_{1}<D\leq D_{2},\\
\text{for }D>D_{2}%
\end{array}
\label{Opt_Offer}%
\end{equation}
where $D_{1}$ and $D_{2}$ are given in Eqs.~(\ref{D1},~\ref{D2}), as obtained
by Selten. FIG.~2 plots the optimal offer against the demand $D$ when the
probability of capturing the kidnapper depends on whether the hostage has been
executed or not (dotted line) and in the case studied by Selten (solid line).
Note that, with reference to Eqs.~(\ref{D1},~\ref{D2},~\ref{D1_New}%
,~\ref{D2_New}), the figure assumes that $D_{2}>D_{1}$ and $D_{2}^{\prime
}>D_{1}^{\prime}$, but generally $D_{1}^{\prime}-D_{1}\neq D_{2}^{\prime
}-D_{2}$ and $D_{2}-D_{1}\neq D_{2}^{\prime}-D_{1}.$%

\begin{figure}[ptb]%
\centering
\includegraphics[
trim=0.474612in 3.546050in 0.737550in 2.493591in,
height=4.9061in,
width=6.0883in
]%
{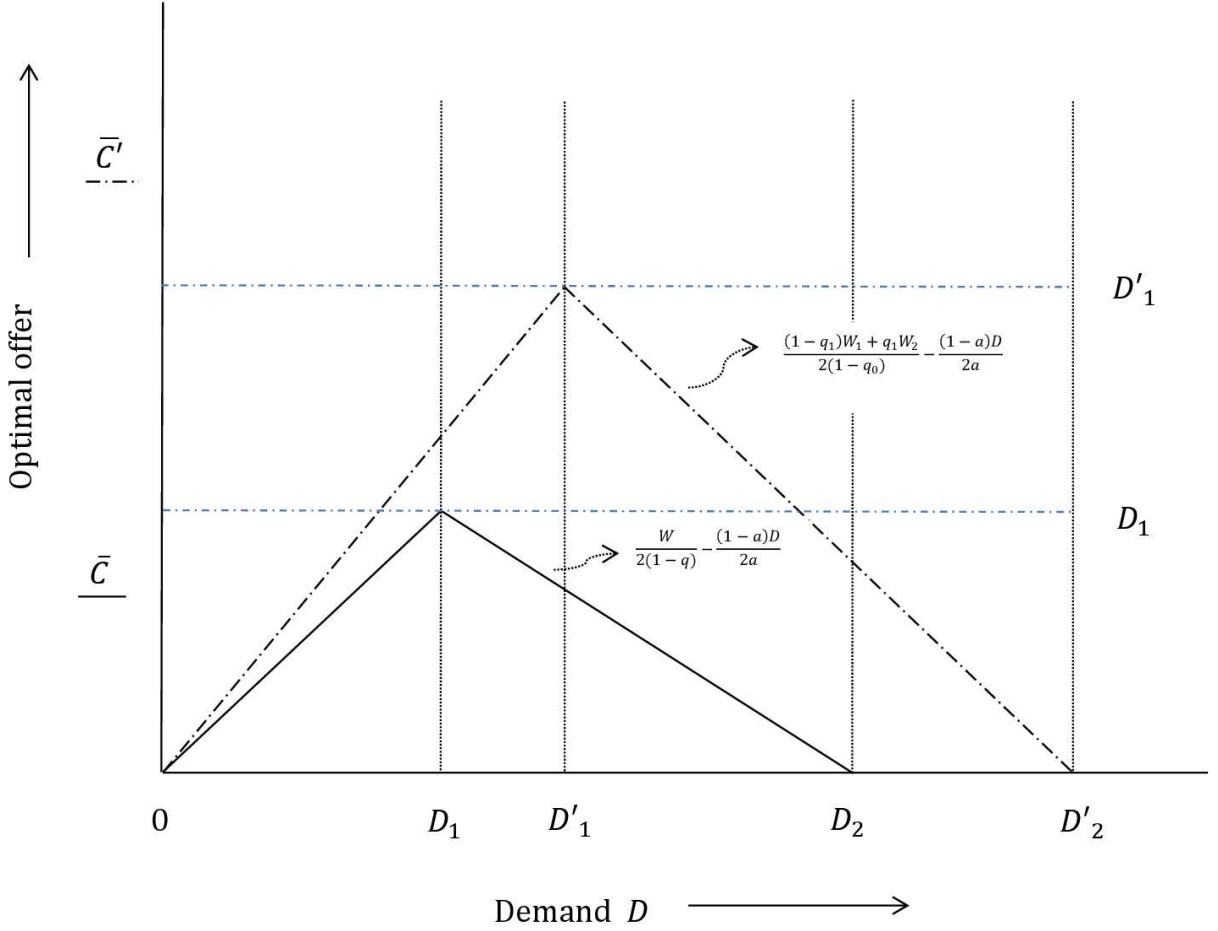}%
\caption{Optimal offer againts the demand when the probability of capturing
the kidnapper depends on whether the hostage has been executed or not (dotted
line) and in the case studied by Selten (solid line). The scales along the
optimal offer and demand axes are different.}%
\end{figure}

\subsection{Optimal Choice of $D$}

We now consider the subgame that begins with player K's choice of $D$, the
amount requested for the ransom. The optimal offer is given by
Eq.~(\ref{Opt_Offer_New}). Let $\bar{\alpha}$ and $\bar{V}_{0}$ be the values
that $\alpha$ and $V_{0}$ assume at $C=\bar{C}$, respectively. Then, the
optimal probability of the non-rational execution of the hostage as a function
of demand $D$ becomes:%

\begin{equation}
\bar{\alpha}^{\prime}=a(1-\frac{\bar{C}^{\prime}}{D}). \label{Alpha_bar}%
\end{equation}

Using Eq.~(\ref{Opt_Choice_e_eq1}), we have:

%

\[
\bar{V}_{0}=(1-q_{0})\bar{C}^{\prime}-q_{0}X.
\]
Therefore, player K's expected payoff becomes:%

\begin{equation}
V=(1-\bar{\alpha}^{\prime})\bar{V}_{0}+\bar{\alpha}^{\prime}V_{1},
\end{equation}
where
\begin{equation}
\bar{V}_{0}=(1-q_{0})\bar{C}^{\prime}-q_{0}X.
\end{equation}
and $V_{1}$ is given by Eq.~(\ref{Opt_Choice_e_eq2}). Therefore:%

\begin{equation}
V=a(1-\frac{\bar{C}^{\prime}}{D})(V_{1}-\bar{V}_{0})+\bar{V}_{0}.
\label{V_New}%
\end{equation}
Using Eq.~(\ref{Opt_Offer_New}), $V$ becomes:%

\begin{equation}
V=\left\{
\begin{array}
[c]{c}%
\bar{V}_{0}\\
\\
a\left[  1-\frac{(1-q_{1})W_{1}+q_{1}W_{2}}{2(1-q_{0})D}+\frac{(1-a)}%
{2a}\right]  (V_{1}-\bar{V}_{0})+\bar{V}_{0}\\
\\
a(V_{1}-\bar{V}_{0})+\bar{V}_{0}%
\end{array}
\right.
\begin{array}
[c]{l}%
\text{for }0<D\leq D_{1}^{\prime},\\
\\
\text{for }D_{1}^{\prime}<D\leq D_{2}^{\prime},\\
\\
\text{for }D>D_{2}^{\prime}%
\end{array}
\label{V_New1}%
\end{equation}
which can also be re-written as:%

\begin{equation}
V=\left\{
\begin{array}
[c]{c}%
(1-q_{0})D-q_{0}X\\
\\
\frac{1}{2}\left[  (1+a)-\frac{a\left[  (1-q_{1})W_{1}+q_{1}W_{2}\right]
}{(1-q_{0})D}\right]  V_{1}+\left[  \frac{(1-a)}{2}+\frac{a\left[
(1-q_{1})W_{1}+q_{1}W_{2}\right]  }{(1-q_{0})D}\right]  \bar{V}_{0}\\
\\
a\left[  (1-q_{1})(-Y)+q_{1}(-Z)\right]  +(1-a)(-q_{0}X)
\end{array}
\right.
\begin{array}
[c]{l}%
\text{for }0<D\leq D_{1}^{\prime},\\
\\
\text{for }D_{1}^{\prime}<D\leq D_{2}^{\prime},\\
\\
\text{for }D>D_{2}^{\prime}.
\end{array}
\label{V_New2}%
\end{equation}

Consider first $D_{1}^{\prime}<D\leq D_{2}^{\prime}$. From Eq.~(\ref{V_New2}),
note that $V$ is a decreasing function of $D$ if $\bar{V}_{0}$ is a constant.
But $\bar{V}_{0}=(1-q_{0})\bar{C}^{\prime}-q_{0}X$ and for $D_{1}^{\prime
}<D\leq D_{2}^{\prime}$ we have $\bar{C}^{\prime}=\frac{(1-q_{1})W_{1}%
+q_{1}W_{2}}{2(1-q_{0})}-\frac{(1-a)D}{2a}$. An increase of $D$ decreases
$\bar{C}^{\prime}$ and thus $\bar{V}_{0}$ is decreased too. That is, $V$ will
be decreased further (relative to the case when $\bar{V}_{0}$ is assumed
constant) when $D$ increases within the interval $D_{1}^{\prime}<D\leq
D_{2}^{\prime}$. So that $V$ as a function of $D$ first increases for values
of $D$ up to $D_{1}^{\prime}$. It then decreases for values of $D$ up to
$D_{2}^{\prime}$ and then remains constant. Recall that $D_{1}^{\prime}$ and
$D_{2}^{\prime}$ are given in Eqs.~(\ref{D1_New},~\ref{D2_New}).

Note that under the constraints~(\ref{Constraints}), Eq.~(\ref{V_New2}) is
reduced to%

\begin{equation}
V=\left\{
\begin{array}
[c]{c}%
(1-q)D-qX\\
\\
\frac{1}{2}\left[  (1+a)-\frac{aW}{(1-q)D}\right]  V_{1}+\left[  \frac
{(1-a)}{2}+\frac{aW}{(1-q)D}\right]  \bar{V}_{0}\\
\\
a\left[  (1-q)(-Y)+q(-Z)\right]  +(1-a)(-qX)
\end{array}
\right.
\begin{array}
[c]{l}%
\text{for }0<D\leq D_{1},\\
\\
\text{for }D_{1}<D\leq D_{2},\\
\\
\text{for }D>D_{2},
\end{array}
\label{V_New0}%
\end{equation}
as discussed by Selten, which is an increasing function for $0<D\leq
D_{1}^{\prime}$. For $D_{1}^{\prime}<D\leq D_{2}^{\prime}$ in (\ref{V_New0}),
as is the case for the function (\ref{V_New2}), if $\bar{V}_{0}$ is a constant
then $V$ is a decreasing function of $D$. But now $\bar{V}_{0}=(1-q)\bar
{C}-qX$ as $\bar{C}^{\prime}$ reduce to $\bar{C}$, given by (\ref{Opt_Offer}),
under constraints~(\ref{Constraints}). An increase of $D$ decreases $\bar{C}$
and thus $\bar{V}_{0}$ is decreased too. That is, $V$ will be decreased
further (relative to the case when $\bar{V}_{0}$ is assumed constant) when $D$
increases within the interval $D_{1}^{\prime}<D\leq D_{2}^{\prime}$.

Player K's optimal demand $\bar{D}^{\prime}$ can be considered as the highest
demand $D_{1}^{\prime}$ such that player F's optimal offer $\bar{C}^{\prime}$
coincides with the demand. To determine $\bar{D}^{\prime}$ we refer to
Eq.~(\ref{Opt_Offer_New}) and set%

\begin{equation}
\bar{C}^{\prime}=\bar{D}^{\prime}\text{ and }D=\bar{D}^{\prime},
\label{At_Optimal_Demand}%
\end{equation}
to have%

\begin{equation}
\bar{D}^{\prime}=\frac{(1-q_{1})W_{1}+q_{1}W_{2}}{2(1-q_{0})}-\frac
{(1-a)\bar{D}^{\prime}}{2a},
\end{equation}
which gives%

\begin{equation}
\bar{D}^{\prime}=\frac{a}{(1+a)}.\frac{(1-q_{1})W_{1}+q_{1}W_{2}}{(1-q_{0})}.
\label{Opt_Demand_New}%
\end{equation}

Eq.~(\ref{Opt_Demand_New}), under the constraints (\ref{Constraints}), then
gives the player K's optimal demand $\bar{D}$ as%

\begin{equation}
\bar{D}=\frac{a}{(1+a)}.\frac{W}{(1-q)}, \label{Opt_Demand}%
\end{equation}
as obtained by Selten. Eq.~(\ref{Opt_Demand}) shows that a higher value of $q$
results in an increase in $\bar{D}$. This also shifts $D_{1}$ and $D_{2}$,
given by Eqs.~(\ref{D1},~\ref{D2}), to higher values.

Thus, if the allocation of more resources to K's capture is linked to an
increase in $q$ then this also results in an increase in the optimal demand
$\bar{D}$. Increasing $q$ by allocating higher resources to police, however,
is not an effective policy as it appears to be. This is because with K's
increased probability of capture F's chances to get the ransom money back are
also increased. This results in an increase in F's willingness to pay and thus
to a higher optimal demand.

In our model, if the probability of capture $q_{1}$ is increased, it also
results in an increase in the optimal demand $\bar{D}$. However, since this
increase only concerns $q_{1}$, the likelihood of K to be captured once he
executes the hostage, does not have the perverted effect of increasing F's
willingness to pay.

\subsection{Optimal choice of b}

The binary decision variable $b$ in (\ref{b_Def}) describes player K's choice
whether or not to go ahead with the plan to kidnap. The game ends if K selects
$b=0$ and the hostage is kidnapped if K selects $b=1$. Considering the subgame
which begins with player K's choice of $D,$ the player K's payoff expectation
$V$ is given by Eq.~(\ref{V_New2}). As noted above, $V$ as a function of $D$
is first increasing up to $D_{1}^{\prime}$ and then decreasing up to
$D_{2}^{\prime}$ and then remaining constant. The optimal value $\bar
{D}^{\prime}$ of $D$ is given by Eq.~(\ref{Opt_Demand_New}). As noted before
Eq.~(\ref{At_Optimal_Demand}), $\bar{D}^{\prime}$ is the highest demand
$D_{1}^{\prime}$ such that player F's optimal offer $\bar{C}^{\prime}$
coincides with the demand. Let $\bar{V}^{\prime}$ be the value of $V$ assumed
at $\bar{D}^{\prime}$. From Eq.~(\ref{V_New2}) we have%

\begin{equation}
V=(1-q_{0})D-q_{0}X\text{ \ \ for }0<D\leq D_{1}^{\prime},
\end{equation}
then%

\begin{equation}
\bar{V}^{\prime}=(1-q_{0})\bar{D}^{\prime}-q_{0}X,
\end{equation}
and using Eq.~(\ref{Opt_Demand_New}) this can be written as%

\begin{equation}
\bar{V}^{\prime}=\frac{a\left[  (1-q_{1})W_{1}+q_{1}W_{2}\right]  }%
{(1+a)}-q_{0}X,
\end{equation}
which at $q_{1}=q_{0}=q$ becomes $\bar{V}=\frac{a}{(1+a)}W-qX$. In Selten's
original work, this shows that if the probability of capture $q$ can be
increased by allocating additional resources to the efforts in finding K then
the possibility of decreasing $\bar{V}$ is only limited by the availability of
the resources. In our model, the increase of either probabilities, $q_{1}$ or
$q_{0}$, leads to an overall decrease in K's utility, and the effect very much
depends on the relative values of $X$, $W_{1}$ and $W_{2}$. In particular, an
increase in $q_{1}$ results in the optimal choice of $b$ likely to be $b=0$,
as the value identified in Eq.~(\ref{FinalEq}) decreases [keeping $q_{0}$
constant], and thus the first condition is more likely to be satisfied.
Similarly, if $q_{0}$ increases (keeping $q_{1}$ constant) then the first
condition, i.e. $b=0$ is more likely to hold. If $\Delta W=W_{1}-W_{2}$ is
sufficiently large however then increasing $q_{1}$ appears to be more optimal
in discouraging to select $b=1$.

Now the optimal choice of $\bar{b}$ is obtained by the following requirements%

\begin{align}
\bar{b}^{\prime}  &  =0\text{ for }\bar{V}^{\prime}<0,\nonumber\\
\bar{b}^{\prime}  &  =1\text{ for }\bar{V}^{\prime}>0,
\end{align}
which can be written as%

\begin{equation}
\label{FinalEq}\bar{b}^{\prime}:\left\{
\begin{array}
[c]{c}%
0\\
1
\end{array}
\right.
\begin{array}
[c]{c}%
\text{for }\frac{a\left[  (1-q_{1})W_{1}+q_{1}W_{2}\right]  }{(1+a)}<q_{0}X,\\
\text{for }\frac{a\left[  (1-q_{1})W_{1}+q_{1}W_{2}\right]  }{(1+a)}>q_{0}X,
\end{array}
\end{equation}
and when $W_{1}=W_{2}=W$, and $q_{0}=q_{1}=q,$ it is reduced to%

\begin{equation}
\bar{b}:\left\{
\begin{array}
[c]{c}%
0\\
1
\end{array}
\right.
\begin{array}
[c]{c}%
\text{for }\frac{aW}{(1+a)}<qX,\\
\text{for }\frac{aW}{(1+a)}>qX,
\end{array}
\end{equation}
as obtained by Selten. Player K's choice whether or not to go ahead with the
plan to kidnap now depends on $W_{1},$ $W_{2},$ $q_{0},$ and $q_{1}$.

\section{Discussion}

The dependence of K's probability of being captured on whether he has executed
the hostage or not can be represented as a bifurcation of $q$ (probability of
K's capture in either case of hostage to have been executed or not) into
$q_{1}$ (probability of K's capture when the hostage has been executed) and
$q_{0}$ (probability of K's capture when the hostage has been released after
paying the ransom).


Overall, we show that increasing either $q_{0}$ or $q_{1}$ leads to a reduced
likelihood of kidnapping, provided that $q_{1}>q_{0}$ (otherwise e=0 is not
necessarily the optimal choice of K). We also show that if the kidnapping took
place, releasing the hostage and paying the ransom remains the optimal choice
for K provided the motivation in assigning values to $q_{0}$ and $q_{1}$ takes
into account the heightened sensitivities, i.e.~authorities spend more
resources when the hostage has been executed so as to increase the likelihood
of capturing player K (i.e.~$q_{1}>q_{0}$). Therefore, increasing $q_{1}$ not
only lessen the likelihood of kidnapping, it also ensures that $q_{1}$ stays
above $q_{0}$ and presents the added benefit of lowering the ransom D. This
means that increasing $q_{1}$ appears to be more optimal than increasing
$q_{0}$.



A question that might arise is whether it is in the interest of police to
advertise the increase in resources, or whether F and K even know about it.
From FIG.1, K's rational decision (dictated by $e=1$ or $e=0$) to execute or
release the hostage respectively, is known to F. Even if the police remain
discreet and do not announce that they are investing more (or less or same)
resources in case where $e=1$, the events $e=1$ or $e=0$ themselves appear
sufficient to result in the bifurcation of $q$ into to $q_{0}$ and $q_{1}$.
Furthermore, studying this bifurcation allows us to understand better the
consequences emanating from increasing either $q_{1}$ or $q_{0}$. As we have
seen earlier, increasing $q_{1}$ may result in overall better outcomes.


The non-rational execution of the hostage is a characteristic of Selten's
model that can be explained using a Bayesian approach, i.e.~by considering the
belief K has about F's ability to pay. If K thinks that F can pay but F
decides not to, this can result in K reacting in a non-rational way, as
proposed by Selten. It is anticipated that by introducing beliefs for K,
regarding whether or not F can match his demand, it can lend a further
perspective to the analysis of this game. In particular, this would result in
considering a Bayesian equilibrium instead of subgame perfect equilibrium.

Selten used a binary decision variable $e$ in Eq.~(\ref{Def_e}) in order to
describe the situation that K can execute the hostage while enacting a
non-rational decision. As the hostage may be executed even when the ransom
demand is met, therefore, $\alpha\neq0$ even if $C=D$. An appropriate
probability function to describe the non-rational situation could be when%

\begin{equation}
\alpha=a\left[  1-\frac{C}{D}+\beta\right]  \text{ for }0\leq C\leq D\text{
and }\beta>0.
\end{equation}

A policy objective is to minimize the optimal demand as given in Eq.
(\ref{Opt_Demand_New}). Given fixed resources can be allocated to police to
increase the chances of capturing K, these resources are better spent towards
increasing $q_{1}$.

If K is aware that F cannot meet his demand, then K could either lower his
demand and/or decide whether to execute the hostage on rational grounds. This
rational decision to execute the hostage depends on probabilities $q_{1}$ and
$q_{0}$, and we know that as long as as $q_{1}>q_{0}$, executing the hostage
is not optimal for K. Thus increasing $q_{1}$, as opposed to allocate
resources to $q_{0}$ is again more desirable.

The other situation that could be incorporated in the model is when player K's
cost of preparing the kidnapping is considered non-negligible and player F's
non-monetary disutilities, other than those incurred by the hostage's life,
are however considered negligible. For instance, player F does not attach any
value to the capture of the kidnapper.

\end{document}